\documentclass[manuscript]{aastex}

\shorttitle{The SCUBA-2 ``All-Sky'' Survey}
\shortauthors{Thompson et al.}

\begin{document}

\title{The SCUBA-2 ``All-Sky'' Survey}

\author{M.~A.~Thompson\altaffilmark{1}, 
S.~Serjeant\altaffilmark{2}, 
T.~Jenness\altaffilmark{3}, 
D.~Scott\altaffilmark{4}, 
M.~Ashdown\altaffilmark{5},
C.~Brunt\altaffilmark{11}
H.~Butner\altaffilmark{3},   
E.~Chapin\altaffilmark{4},
A.~C.~Chrysostomou\altaffilmark{1,3},
J.~S.~Clark\altaffilmark{2},
D.~Clements\altaffilmark{5},   
J.~L.~Collett\altaffilmark{1},    
K.~Coppin\altaffilmark{4,6},  
I.~M.~Coulson\altaffilmark{3},
W.~R.~F.~Dent\altaffilmark{7},  
F.~Economou\altaffilmark{3},
A.~Evans\altaffilmark{8},
P.~Friberg\altaffilmark{3}, 
G.~A.~Fuller\altaffilmark{9},
A.~G.~Gibb\altaffilmark{4},  
J.~Greaves\altaffilmark{10},
J.~ Hatchell \altaffilmark{11},
W.~S.~Holland\altaffilmark{7},
M.~Hudson\altaffilmark{12},
R.~J.~Ivison\altaffilmark{7},
A.~Jaffe\altaffilmark{5}, 
G.~Joncas\altaffilmark{22}
H.~R.~A.~Jones\altaffilmark{1}, 
J.~H.~Knapen\altaffilmark{23,1},
J.~Leech\altaffilmark{3},
R.~Mann\altaffilmark{13},
H.~E.~Matthews\altaffilmark{14},
T.~J.~T.~Moore\altaffilmark{15},
A.~Mortier\altaffilmark{13},
M.~Negrello\altaffilmark{2},
D.~Nutter\altaffilmark{16},  
M.~P.~Pestalozzi\altaffilmark{1},
A.~Pope\altaffilmark{4}, 
J.~Richer\altaffilmark{17},
R.~Shipman\altaffilmark{18},
J.~S.~Urquhart\altaffilmark{19},
M.~Vaccari\altaffilmark{5, 24},
L.~Van Waerbeke\altaffilmark{4}, 
S.~Viti\altaffilmark{20},
B.~Weferling\altaffilmark{3},
G.~J.~White \altaffilmark{2, 21},
J.~Wouterloot \altaffilmark{3}, 
M.~Zhu \altaffilmark{3, 14}}

\altaffiltext{1}{Centre for Astrophysics Research, Science \& Technology Research Institute, University of Hertfordshire, College Lane, Hatfield, AL10 9AB, UK}
\altaffiltext{2}{The Open University, Walton Hall, Milton Keynes MK7 6AA, UK}
\altaffiltext{3}{Joint Astronomy Centre, 660 N.~A'ohoku Place, Hilo, HI 96720, US}
\altaffiltext{4}{Department of Physics and Astronomy, University of British Columbia, 6224 Agricultural Road, Vancouver, BC, V6T 1Z1, CA}
\altaffiltext{5}{Imperial College, London, UK}
\altaffiltext{6}{University of Durham}
\altaffiltext{7}{UK Astronomy Technology Centre}
\altaffiltext{8}{Astrophysics Group, Keele University, Keele, Staffordshire, ST5 5BG}
\altaffiltext{9}{School of Physics and Astronomy, University of Manchester, Sackville Street, PO Box 88, Manchester M60 1QD, UK}
\altaffiltext{10}{University of St Andrews, UK}
\altaffiltext{11}{University of Exeter, UK}
\altaffiltext{12}{University of Waterloo, Canada}
\altaffiltext{13}{University of Edinburgh, UK}
\altaffiltext{14}{Herzberg Institute of Astrophysics/National Research Council of Canada, 5071 W. Saanich Rd, Victoria, BC, V9E 2E7, CA}
\altaffiltext{15}{Astrophysics Research Institute, Liverpool John Moores University, Twelve Quays House, Egerton Wharf, Birkenhead, CH41 1LD, UK}
\altaffiltext{16}{Cardiff University, UK}
\altaffiltext{17}{Cavendish Laboratory J J Thomson Avenue Cambridge CB3 0HE, UK}
\altaffiltext{18}{SRON Netherlands Institute for Space Research, PO Box 800, 9700 AV, Groningen, NL}
\altaffiltext{19}{University of Leeds, UK}
\altaffiltext{20}{Department of Physics and Astronomy, University College London, Gower St.,
London, WC1E 6BT, UK}
\altaffiltext{21}{CCLRC Rutherford Appleton Laboratory, Chilton, Didcot Oxfordshire, OX11 9DL, UK}
\altaffiltext{22}{Dept. de physique, de g\'enie physique et
   d'optique, Universit\'e Laval, Qu\'ebec, Qu\'ebec, Canada G1K 7P4}
\altaffiltext{23}{ Instituto de Astrof\'\i sica de Canarias, E-38200 La Laguna, Spain}
\altaffiltext{24}{Department of Astronomy, University of Padova, Italy}

\begin{abstract}  
The sub-millimetre wavelength regime is perhaps the most poorly explored over large areas of the sky, despite
the considerable effort that has been expended in making deep maps over small regions. As a consequence the
properties of the sub-millimetre sky as a whole, and of rare bright objects in particular, 
remains largely unknown.
 Here we describe a forthcoming survey
(the SCUBA-2 ``All-Sky'' Survey, or SASSy) designed to address this issue by making a large-area map
of approximately one-fifth of the sky visible from the JCMT (or $\sim$4\,800 square degrees) down to a
1$\sigma$ noise level of 30 mJy/beam. This map
forms the pilot for a much larger survey, which will potentially map the remaining sky visible from
the JCMT, with the region also visible to ALMA as a priority. SASSy has been awarded 500 hours for the
4\,800 square degree pilot phase and will commence after the commissioning of SCUBA-2, expected in early 2008.  
\end{abstract}



\section{Introduction: The JCMT Legacy Surveys}

The James Clerk Maxwell Telescope (JCMT) is in the process of upgrading its instrument suite to
include the new wide-field bolometer camera SCUBA-2  \citep[the Sub-mm Common-User Bolometer Array 2;
][]{scuba2}. The radically new set of detectors at the heart of SCUBA-2 and its large instantaneous
field of view (50 square arcminintes) gives SCUBA-2 a mapping speed some three orders of magnitude
larger than its predecessor SCUBA. This vast increase in mapping speed will bring about a new era of
wide-field sub-millimetre astronomy and finally brings about the possibility  of deep (tenss of mJy) 
surveys encompassing significant fractions of the sky.
Recognising the unique capabilities of SCUBA-2 and the new heterodyne array HARP-B \citep{harp-b} the
JCMT Board called upon the JCMT community to identify and propose an innovative programme of
large-scale Legacy Surveys. The JCMT Legacy Survey Programme\footnote{Further details of the JCMT
Legacy Survey Programme may be found at \texttt{http://www.jach.hawaii.edu/JCMT/surveys}} has
recently been ratified by the JCMT Board and consists of seven individual survey projects stretching
from an in-depth spectral survey of the 350 GHz band toward a number of star-forming regions and PDRs
\citep[the Spectral Legacy Survey;][]{sls} to a nested series of deep 450 \& 850 $\mu$m cosmology
surveys (the JCMT Cosmology Survey). In this paper we describe perhaps one of the most ambitious JCMT
Legacy Surveys, a project to map a large fraction of the sky: the \textbf{S}CUBA-2
``\textbf{A}ll-\textbf{S}ky'' \textbf{S}urve\textbf{y}, or SASSy.


SASSy is a project designed to use SCUBA-2 to
map roughly one-fifth of the sky visible from the JCMT (4\,800 square degrees) down to an 
r.m.s.~level of 30 mJy at 850 $\mu$m over a two year pilot phase. This pilot phase is intended as a
pathfinder to a much larger area survey, perhaps encompassing the entire sky visible from JCMT
(23\,000 square degrees). We will use the results from the pilot phase to inform the design and
strategy of the later survey. SASSy has been awarded 500 hours in which to carry out this two year pilot
phase and uses a novel approach to carry out the survey in sub-optimal weather (JCMT Grade 4 weather;
$0.12 \le \tau_{\rm 225 GHz} \le 0.2$). This allows us not only to carry out such a large-area survey
at minimal cost to the telescope programme but to also provide an ideal poor-weather backup project,
enhancing the productivity of the JCMT's  flexible scheduling. The SASSy Consortium includes over forty astronomers
from the UK, Canada \& the Netherlands and is led by four coordinators representing the galactic \&
extragalactic science communities: Mark Thompson, Stephen Serjeant, Tim Jenness \& Douglas Scott.
SASSy will commence with the rest of the JCMT Legacy Survey programme immediately subsequent to
SCUBA-2 commissioning on the JCMT, which is expected to be completed in early 2008.

\subsection{The need for an all-sky sub-mm survey}

The submillimetre sky remains the most poorly surveyed to date, in contrast to the very successful IRAS
\citep{beichmann1988}, 2MASS \citep{cutri2003} and Digitised Sky Survey (DSS) surveys of the
mid/far-infrared, near-infrared and visible-wavelength sky. IRAS in particular has proved to be a
valuable resource in fields as diverse as solar system studies, star formation and 
interacting galaxies. The legacy value of these large-scale surveys cannot be understated: even now,
over 20 years after the IRAS mission ended, IRAS data are still in active use
\citep[e.g.][]{hwang2007,garay2007,lowe2007},  and IRAS has generated $>40,000$ citations in refereed
journals alone to date. However, whilst IRAS was responsible for ushering in a new era in infrared
astronomy, its main limiting factor was the relatively small size of its primary mirror, which imposed
the twin problems of coarse angular resolution (typically $\sim$3$\times$5\arcmin\ at 100$\mu$m) and limited
sensitivity. The strong desire of the astronomical community to overcome these issues can be evidenced
in the development of new IRAS resolution enhancement techniques (HIRES, \citealt{cao1997}; IRIS,
\citealt{miville2005}, all-sky dust maps \citealt{schlegel1998} --- one of the most highly cited papers
in astrophysics), new space telescopes with higher resolution and sensitivity (e.g.~ISO,
Spitzer and Herschel), and new all-sky survey missions in the form of Akari and WISE.

The four infrared bands surveyed by IRAS opened the window upon the warm and luminous Universe
of comets, debris disks, young stellar objects and ultraluminous starbursts.  But in the
post-IRAS era the development of ground-based millimetre and sub-millimetre astronomy, and
in particular the deployment of  bolometer arrays like SCUBA, began to reveal a much
colder, darker Universe of prestellar and starless cores, infrared-dark molecular clouds,
cold dust in nearby galaxies and high-redshift sub-mm galaxies that were either too cold or
too faint to be detected by IRAS. Only a miniscule fraction of the submillimetre sky has
been explored to any level of detail: although the COBE mission made an all-sky sub-mm map,
it did so with only 7\degr\ angular resolution \citep{fixsen1996}. The Planck Surveyor satellite
\citep{tauber2004} will extend the IRAS infrared coverage to the sub-mm wavelength range.
However, Planck's angular resolution of 5\arcmin\ in the sub-mm wavelength bands will not
improve upon the angular resolution of IRAS and this low angular resolution will subject
Planck data to considerable confusion, both within the crowded Galactic plane and to
considerable distances away from the plane (here due to cirrus confusion). This not only
limits the production of a point source catalogue from Planck data but it is also a factor
in the subtraction of the galactic foreground from the CMB measurements which is Planck's
primary goal. So whilst Planck will provide data of exquisite sensitivity and wavelength
coverage over the entire sky, the low angular resolution of the satellite limits the study
of galactic and extragalactic sources to regions of relatively low cirrus and
confusion noise. 

The fact that Planck and the Herschel Space Observatory \citep{pilbratt2004} share the same launch
vehicle brings about a third area of difficulty. Both missions will begin science operations at
approximately the same time, implying that Planck will not be able to act in the same pathfinding
role for Herschel that the IRAS satellite did for ISO. The Planck Science Team are investing
considerable effort into producing an Early Release Compact Source Catalogue (ERCSC) some 22--24 months
post-launch in order to meet the final Herschel Announcement of Opportunity and aid the Herschel
community in planning final Herschel observations. However, the ERCSC will be shallower than the
final catalogue by approximately an order of magnitude. Only the brightest sources found by Planck
will be available for Herschel follow-up before the cryogen supply on Herschel is depleted and the
mission is over.

There is thus a pressing need for a sub-millimetre sky survey that is both timely (to capitalise upon
Herschel's limited lifetime) and possesses high angular resolution (to address problems of source
confusion and cirrus noise). The sub-millimetre wavelength regime acts as  an optically thin tracer of
cold dust and as a  redshifted tracer of far-infrared emission from distant high-luminosity galaxies
and a large area unbiased sub-mm survey will revolutionise many fields in galactic and extragalactic
astronomy from the formation of massive stars to the number counts of furiously star-forming galaxies. 
Such a timely survey has many legacy benefits to a number of planned facilities beyond Herschel; it
will act as a sub-mm finding chart for the ALMA interferometer, will provide formidable aid to many of
the science goals of Planck by virtue of allowing high resolution foreground subtraction, and will
supply lists of potential pointing and flux calibrators to millimetre \& sub-millimetre facilities
worldwide. SASSy and SCUBA-2 offer the unique opportunity to achieve these goals and add considerable
value to some of the most important new facilities in sub-millimetre \& far-infrared astronomy
(Herschel, ALMA \& Planck). 


\section{Survey Strategy}

Even though SCUBA-2 has a mapping power a thousand times greater than that of SCUBA, a considerable
amount of time is still required to map sky areas of several thousands of square degrees. One of the
major difficulties in designing SASSy was to minimise the impact of such a large-scale survey on the
demands for telescope time (particularly in the era of large-scale Legacy Surveys) whilst maximising
the potential science benefits that the survey could bring. It was decided that the central role of
SASSy was to be a pure detection experiment aimed at relatively bright galactic and extragalactic
sources. By restricting SASSy to 850 $\mu$m observations alone, it becomes possible to use the
capabilities of SCUBA-2 to work in less favourable weather conditions than the traditional weather
bands usually reserved for SCUBA, which not only increases the observing efficiency of the facility
by increasing the amount of available observing time, but also provides a poor weather fallback
project against which other observations can be flexibly scheduled.

There are four new qualities of SCUBA-2 which make this possible: \emph{i)} the SCUBA-2 detectors
sample data at 200 Hz, which is faster than the ``knee'' in the atmospheric power spectrum \emph{ii)}
SCUBA-2 has many more pixels than SCUBA, which will improve the cancellation of sky noise;
\emph{iii)} the SCUBA-2 pixels are inherently more sensitive than those of SCUBA (by around a factor
of 3--5); and \emph{iv)} the total power operation of SCUBA-2 removes the need for multiple scans to 
remove chopping artefacts. All in all these features combine to make  Grade 4 SCUBA-2 850 $\mu$m operations roughly
equivalent to SCUBA in Grade 2 weather.  Working in Grade 4 weather conditions  does not compromise the science goals of SASSy; the forecast sensitivity and
mapping speed of SCUBA-2  and scan-rate means that it is possible to reach a 1$\sigma$ noise of 30
mJy/beam over the survey area within a reasonable period. 

Using a conservatively scaled  Noise Equivalent Flux Density (NEFD) of 50 mJy Hz$^{-1/2}$ for $\tau_{\rm
225 GHz} = 0.2$ and an observing efficiency of 60\%  we estimate that we will be able to map an area of
10 square degrees to an r.m.s.~noise level of 30 mJy/beam in one hour of observing time. This
sensitivity level is desirable for a number of reasons: it is directly comparable to the end-of-mission
sensitivity of the 850 $\mu$m channel of Planck Surveyor \citep{tauber2004}; it is similar to many of 
the earlier unbiased star formation surveys with SCUBA \citep[e.g.][]{mitchell2001,
johnstone2004,hatchell2005};  5\,$\sigma$ detections from this flux level (i.e. 150 mJy/beam) are at the
level at which the number counts of sub-mm galaxies are at their most uncertain (see
Sect.~\ref{sect:scigoals} and \ref{fig:galaxy_counts}) and at which the number of gravitationally lensed
sources peaks \citep[according to the model of][]{perrotta2003}; and it is sufficient to detect
solar-mass star-forming cores out to a distance of 1 kpc and massive star-forming cores ($\sim$400
M$_{\odot}$) to the far edge of the Galaxy. This sensitivity level is also a natural choice in terms of
likely observing modes as it represents the noise level obtained by mapping a given region three times
at the nominal scan rate of SCUBA-2. 

We
will operate SASSy in fully flexibly-scheduled mode to optimise the efficiency of the
survey, i.e.~target fields will be selected based purely upon the current weather
conditions (within our allocated band $0.12 \le \tau_{\rm 225 GHz} \le 0.2$) and elevation.
We will also use the full width of our weather band to ensure an even survey depth:
fields at high airmass will be mapped at low values of $\tau_{\rm 225 GHz}$ and vice versa.

\subsection{2 year pilot phase}

In common with many of the other JCMT Legacy Surveys SASSy is split into a two year pilot
phase and a follow-on phase. During its 2 year pilot phase SASSy will concentrate upon 
mapping two 10\degr\-wide
strips of sky encompassing 4850 square degrees, or roughly a fifth of the sky visible to the
JCMT. Both strips will be mapped to the same depth of 30 mJy/beam. Each of these strips in
shown in Figure \ref{fig:2year} with a scale representation of the 25 square degree area
that SCUBA mapped in its 8 year lifetime. The average depth of the entire 25 square degree
SCUBA archive is $\sim$ 30 mJy/beam (Johnstone, priv.~comm.) and so the pilot phase of SASSy
may be thought of as compiling the equivalent of 194 SCUBA archives in just 2 years. The
pilot phase allows us to complete a number of the science goals of SASSy (see
Sect.~\ref{sect:scigoals}) and will provide early self-contained source catalogues from each
mapped strip. 

The first 10\degr\-wide strip is centred on the Galactic plane, running from $0\degr \le l \le
245\degr$ and with $|b|\le 5$. This strip is informally known as GP-Wide and is perfectly matched in
latitude to the MSX mid-infrared \citep{price2001} and UKIDSS near-infrared Galactic Plane Surveys
\citep{warren2007}. The second strip is perpendicular to the Galactic Plane and takes the form of a
great circle passing through the North Ecliptic Pole (at $l = 96\degr$), North Galactic Pole and South
Galactic Pole. This strip is informally known as Pole-to-Pole and is truncated at both ends by the
declination limit of the survey ($\delta\ge -30\degr$). As SASSy will be observed in Grade 4 weather we
have chosen stringent declination limits to avoid observing low-lying target fields with too large an
airmass.  The Pole-to-Pole strip is chosen to pass over the North Ecliptic Pole as this is where the
Planck Surveyor will obtain one of its deepest survey regions \citep{tauber2004}. In addition, a number
of high-latitude cloud complexes are known to lie within the Pole-to-Pole strip, including a portion of
the largest and most molecule-rich complex, MBM 53--55 \citep{chastain06}, and the 
Pegasus and Lacertae high-latitude clouds \citep{dame2001}.

SASSy has been awarded 500 hours of Grade 4 weather to complete this 2 year pilot phase, which
should commence with the successful commissioning of SCUBA-2 on the JCMT in early 2008. 

\begin{figure*}[] 
\caption{Two views of the IRAS
100 $\mu$m  sky survey, showing the areas to be mapped during the  2 year pilot phase of SASSy. Each
strip is 10\degr\ wide and is terminated by the planned declination limit of the survey ($\delta \ge
-30\degr$). The GP-Wide is  centred on the Galactic Plane running from $0\le l\le 245$ and the
Pole-to-Pole strip is centred on the North Ecliptic Pole ($l$ = 96\degr). For comparison, the black
central square shows to approximate scale the 25 square degree  area that was mapped by SCUBA in its
entire 8 year lifetime.} 
\label{fig:2year}
\end{figure*}

\subsection{The ``All-Sky'' Survey phase}

\begin{figure*}[]
\caption{The region of sky to be mapped in the ``all-sky'' survey plan. Shaded areas indicate the survey
regions that will be mapped during the 2 year pilot phase. Following the completion of the 2 year
pilot  we will concentrate
upon the band of sky that is visible to both JCMT and ALMA ($-30\degr \le \delta \le +40\degr$), which
is indicated by the lower two solid lines. After this phase is completed we will complete the northern
cap ($+40 \le \delta \le +70\degr$).}  
\label{fig:5year}
\end{figure*}

We will use the results from the 2 year pilot survey to inform our later strategy for a larger area
survey. The source counts of bright sub-mm galaxies and isolated protostars are currently extremely
uncertain and the data from the pilot survey are needed to plan the most effective followup survey,
e.g.~whether to go deeper or shallower, or to prioritise areas such as low galactic latitudes, etc.
One option for the following survey (shown in Fig.~\ref{fig:5year}) is to cover the remaining sky
visible from the JCMT to the same target depth as the pilot survey, i.e. to a depth of 30 mJy,
focusing first on the region of sky visible to both ALMA and the JCMT (an equatorial band with
$-30\degr \le \delta \le +40\degr$), then completing the northern cap which is not visible from ALMA
($+40\degr \le \delta \le +70\degr$).  The extreme northern latitudes ($\delta \ge 70\degr$) lie outside our  elevation
limits for the survey, although if it becomes desirable to complete the survey to high
latitude it is possible to achieve this at lower elevation in better weather conditions.
This strategy allows us to focus our attention upon those
regions of sky which offer the most legacy benefit to future facilities such as ALMA and CCAT, whilst
maintaining a sufficient survey depth to achieve the unbiased survey of star formation that is the
central goal of SASSy.


\section{Science Goals of SASSy}
\label{sect:scigoals}

The main science goal of SASSy is to perform an unbiased search for star formation across the sky,
whether the star formation is located in massive Infrared Dark Clouds (IRDCs), in known molecular
clouds, isolated from known regions, or in furiously star-forming galaxies at high redshift. Our
secondary goal is to search for emission from cold dust that was missed by IRAS, either due to its low
sensitivity to cold dust or its high level of confusion, and hence to reveal the
presence of as yet unknown cold molecular cloud cores in our Galaxy and cold dust in nearby galaxies.
To achieve these goals we will produce a sensitive large-area unbiased 850 $\mu$m map of a large
fraction of the sky (ultimately all of the sky visible from the JCMT), exploiting the two properties of
850 $\mu$m emission as both an optically thin mass tracer in the local Universe and as a redshifted
tracer of the strong far-infrared emission from distant ultraluminous galaxies. The organisation and
science working groups of SASSy are split into galactic and extragalactic science themes and we outline
the science goals of each theme below.

\subsection{Galactic Science Goals}

The main Galactic science goal of SASSy is  simply to answer the question: where do stars
form in our Galaxy? The major sites of star formation as revealed by IRAS are dark and
giant molecular clouds, but in more recent years it has become apparent that stars are
forming in a range of environments unseen by IRAS, e.g.~IRDCs
\citep{carey2000}, high galactic latitude molecular "cirrus" clouds \citep{heithausen2002}, 
and isolated regions far from known star-forming complexes (e.g.~the distributed T-Tauri
problem of \citealt{feigelson1996}). One recently discovered and telling  example is the
isolated candidate class I protostar, 2MASS 0347392+311912, found by \cite{young2006}
during the Spitzer Legacy Survey \emph{Cores to Disks}. This object was discovered
accidentally during ancillary  SCUBA observations of Perseus molecular cloud cores when the
coordinates of a core within Perseus were incorrectly entered. 2MASS 0347392+311912 lies
some 30\arcmin\ from the bulk of the molecular gas in the Perseus East complex (as shown in
Fig.~\ref{fig:perseus}) but has an SED characteristic of a class I protostar with an 850
$\mu$m flux of over 0.2 Jy \citep{young2006}. However, 2MASS 0347392+311912 is but one
example of a serendipitous discovery; how many more isolated examples of star formation are
there in our Galaxy and what is the fraction of stars that form in such isolated
environments?

\begin{figure*}
\caption{The Eastern half of the Perseus molecular cloud in $^{13}$CO J=1--0 emission,
taken by the COMPLETE survey \citep{ridge2006}. The circle indicates the position of the SCUBA
detection of 2MASS 0347392+311912 by \citet{young2006}.}
\label{fig:perseus}
\end{figure*}

SASSy aims to answer these questions by providing a relatively unbiased volume-limited
sample of star forming cores (like 2MASS 0347392+311912), in the sense that 850 $\mu$m continuum emission is sensitive
to both the warm and cold proto- and prestellar dust cores enshrouding the early stages of
star formation. Such a survey will  reveal the whole panoply of Galactic star formation from
isolated star-forming regions, to diffuse high latitude clouds and giant molecular clouds.
Comprehending star formation in all its environments is crucial to our understanding of the
formation of molecular clouds and their evolution throughout the galactic ecology. We will
obtain a true  volume-limited sample only with the results from the ``All-Sky'' survey, but
in the pilot phase we will begin by compiling censuses of IRDCs
found in the GP-Wide strip and high-latitude star formation in the Pole-to-Pole strip.

IRDCs were first found as patches of extinction in the mid-infrared MSX Galactic Plane
survey \citep{egan1998}. Since their discovery a number of IRDCs have been shown to be
dense, cold, gravitationally bound molecular cloud cores that may be intimately involved
with the earliest stages of massive star formation (\citealt{price2002};
\citealt{pillai2006};  \citealt{rathborne2006}). Over 10\,000 candidate IRDCs have been
identified in the MSX Galactic Plane survey \citep{simon2006}, but because these objects
have only been observed in extinction against the Galactic Plane we cannot be certain of
their true distribution. In order to be detected in mid-IR extinction the IRDCs must be
highly contrasted against the diffuse emission from the Galactic Plane. IRDCs that are
distant (with most of the diffuse emission along the line of sight in the foreground) or
IRDCs located in the Outer Galaxy (where the diffuse emission is weak) were not detected by
MSX. This effect can be seen in Fig.~7 of \citet{simon2006},  where the distribution of
IRDCs is seen to clearly follow that of the diffuse Galactic emission in both galactic
latitude and longitude. Indeed, less than 1\% of the known IRDCs are located in the Outer
Galaxy. An uncertain fraction of  the candidate IRDCs found by \citet{simon2006}
may simply be voids in the complex small-scale emission of the Galactic Plane.

Obtaining a reliable catalogue of IRDCs which reveals their true distribution in the
Galaxy, their physical properties and their star-forming nature is crucial to understanding
the relationship of these clouds to the star formation cycle and to Galactic evolution as a
whole. IRDCs are known to emit strongly in sub-mm continuum \citep{rathborne2006} and as
the GP-Wide strip of SASSy is perfectly matched in latitude to the MSX Galactic Plane
survey strip ($|b|\le 5\degr$) we will be able to compile such a catalogue of IRDCs by
comparing the SASSy 850 $\mu$m images with the MSX Galactic Plane Survey mid-IR images.

In the Pole-to-Pole strip we will investigate the star-forming potential of high-latitude
clouds by performing an unbiased search for protostellar cores and sub-Jeans mass cores
within and near these clouds. A number of known high-latitude clouds lie within the
Pole-to-Pole strip, including the eastern portion of the MBM53--55 complex, the Lacertae
and G272.9+29.3 clouds, and the MBM26, 32--33, 43 and 44 clouds. These high-latitude clouds
form a distinctly local population approximately 200 pc distant and the sensitivity limits
of SASSy are such that we will be sensitive to sub-Jeans mass cores and circumstellar discs
such as those found in MBM12 by \citet{hogerheijde2003}. SASSy will thus perform a
sensitive search of these clouds for dust emission from both embedded cores and young
stellar objects. We will correlate the  positions of detected objects in SASSy against the
known high latitude clouds from \citet{magnani1996} and \citet{dobashi2005} to determine
whether there is indeed an unknown population of star formation in high-latitude cirrus
clouds or small molecular cloudlets.

\subsection{Extragalactic Science Goals}

The main extragalactic science goals of SASSy are to discover new and extreme populations
of rare galaxies and to provide high-resolution maps of Galactic foregrounds as an input to
Planck Surveyor. By virtue of the extremely large area that will be surveyed, SASSy has a
tremendous potential for detecting objects that are too rare to have been found in the
current and planned deep pencil-beam extragalactic surveys (e.g. the 8mJy survey,
\citealt{scott2002}; SHADES, \citealt{mortier2005}). These rare sources include extreme
luminosity galaxies at high redshift, strong gravitational lenses and potential cold local
ultraluminous galaxies. The depth of SASSy is carefully optimised to detect extragalactic
populations that are inaccessible to the JCMT Cosmology Legacy Survey. Between 0.2 and 1
source (\citealt{rr2001}; \citealt{pearson2005} respectively) at a flux level of 150 mJy is
expected to be found in the 20 square degree JCMT Cosmology Survey, whereas in the 130
times larger Pole-to-Pole strip we would expect to find between 30 and 130 sources for
these two representative models. SASSy is thus a valuable adjunct to much deeper narrower
surveys and the ensemble of surveys makes it possible to sample the complete range of
sub-millimetre galaxy properties.

The 2300 square~degrees of the
Pole-to-Pole
strip at $z>0.5$ comprises a comoving volume some six times that of the entire $z<0.5$
Universe. For the sky visible from JCMT this rises to some 54 times the $z<0.5$ comoving
volume. With SASSy we will therefore have the remarkable potential to find galaxies that
are too rare to have \emph{any} local counterparts.By sampling these extreme objects, the
properties and trends which are only weakly  apparent in the existing SCUBA population will
become much more apparent. We will confirm the high redshift of detected objects using a
new photometric redshift estimator; objects detected by SASSy but undetected by the Akari
survey must be high-redshift ($z\stackrel{>}{_\sim}2$) hyperluminous systems.  The
high-luminosity galaxies  detected by SASSy will also be much easier follow-up targets than
other SCUBA/SCUBA-2 galaxy populations,  e.g.~to determine redshifts without recourse to
optical IDs via millimetre-wave CO lines (providing a bright target  list for commissioning
the ALMA CO redshift technique). 

\begin{figure*}
\begin{center}
\vspace*{-0.5cm}\hspace*{-1.5cm}\includegraphics*[width=10cm,trim=0 30 0 0]{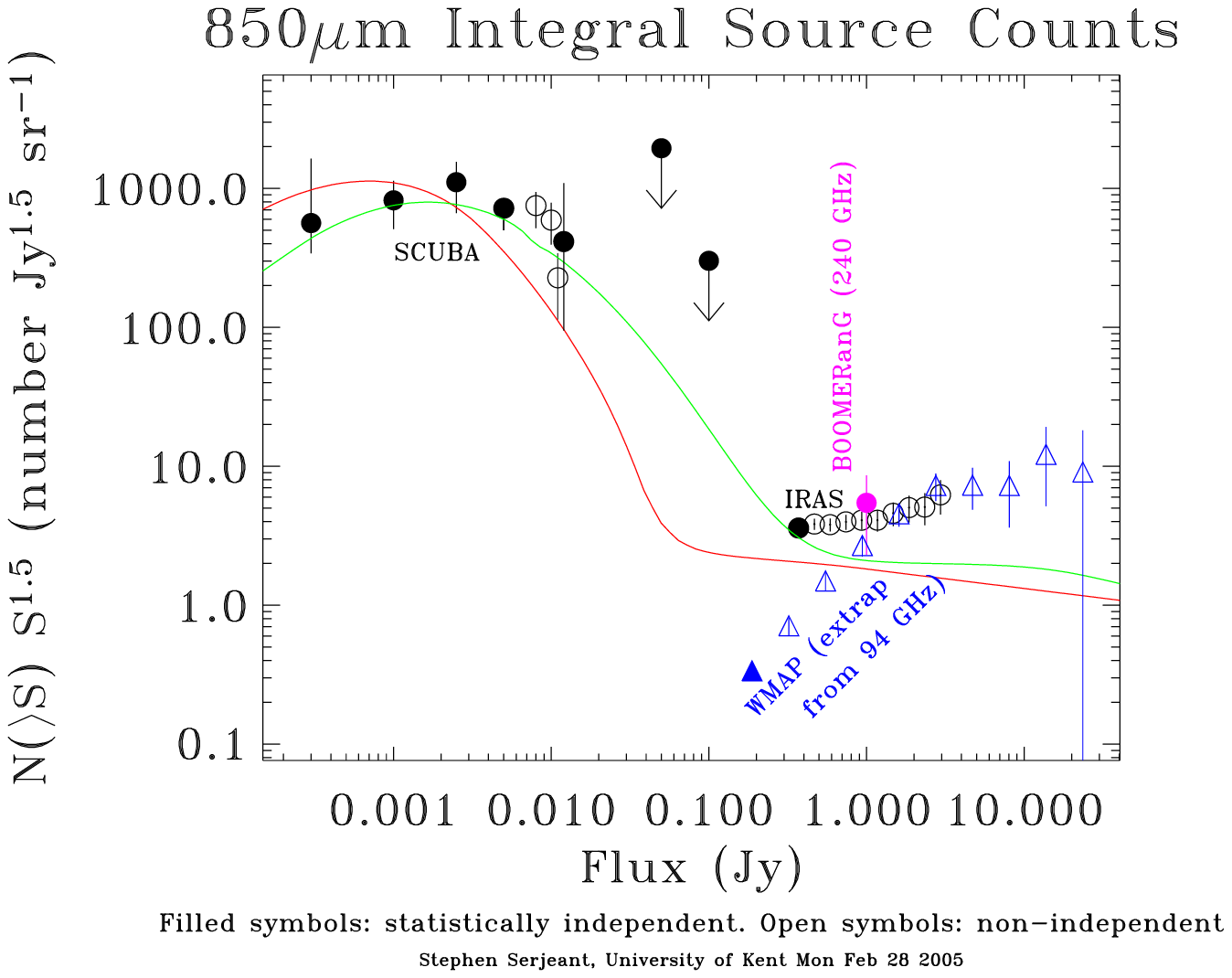}
\hspace*{-3cm}\includegraphics*[width=10cm,trim=0 30 0 0]{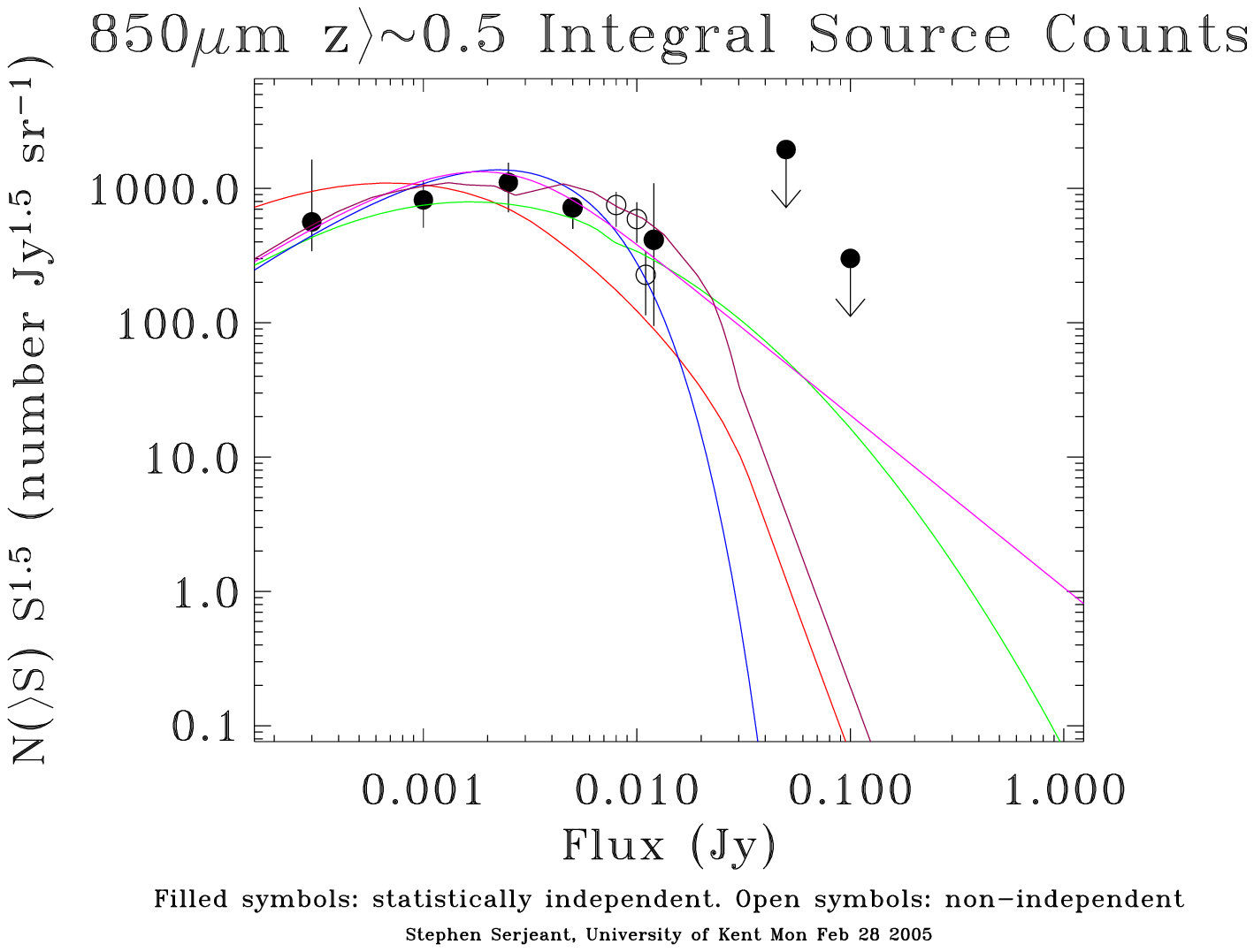}
\caption{\label{fig:galaxy_counts}
{\bf (Left)} Integral source counts, normalised to the Euclidean slope.
The red curve is the prediction from
\citet{rr2001}, and the green curve is from \citet{pearson2005}. 
{\bf (Right)}, as left, except the Pearson and
Rowan-Robinson models have been restricted to $z>$0.5, and in
addition are overplotted: a Schechter function fit (blue) to the
high-$z$ counts; a broken power-law fit (magenta); and the \citet{granato2001}
 model (brown, extrapolated from published curve at
fluxes $>$30mJy). 
}
\end{center}
\end{figure*}

The right-hand panel of Fig.~\ref{fig:galaxy_counts} shows the current constraints on
the sub-mm number counts for high-redshift objects, compared to several source count
models (\citealt{rr2001}; \citealt{pearson2005}; \citealt{granato2001}) which have been
constrained to reproduce the observed $<$10 mJy counts, and reproduce them very well.
However, at higher fluxes they diverge wildly.  The left-hand panel of Fig.\
\ref{fig:galaxy_counts} shows the Pearson and Rowan-Robinson models plotted against the
observational constraints of high-redshift SCUBA galaxy surveys (\citealt{scott2002};
\citealt{cowie2002}; \citealt{barger1999}; \citealt{webb2003}; \citealt{barnard2004}); 
local galaxy source counts derived by \citet{serjeant2005}; an
estimate of the radio-loud AGN source counts derived from SED fits to the objects
listed in the WMAP point source catalogue; and an estimate of the radio source counts
derived from BOOMERANG. Local galaxies are distributed with a Euclidean slope, i.e.\
along a horizontal line, while deviations above the horizontal line are indications of
high-redshift, high-luminosity populations. There is a substantial gap between the
high-redshift and low-redshift populations (at $S_{850} \sim$0.1 Jy) which no previous
survey has been able to constrain.

The main physical differences between the models lie in the
assumptions on the feedback processes and IMFs. Even at the
$\sim10$ mJy level, these processes are parameterised in an ``ad hoc''
and semi-empirical way. SASSy will be the first survey to
determine whether feedback processes provide an upper limit to the star
formation rate in a galaxy and the first survey to uncover the source
counts in the 150~mJy regime on the approach to the Euclidean slope,
that is missing from both panels in Fig.\ \ref{fig:galaxy_counts}.

\begin{figure} 
\includegraphics*[width=\linewidth,trim=20 0 50 0]{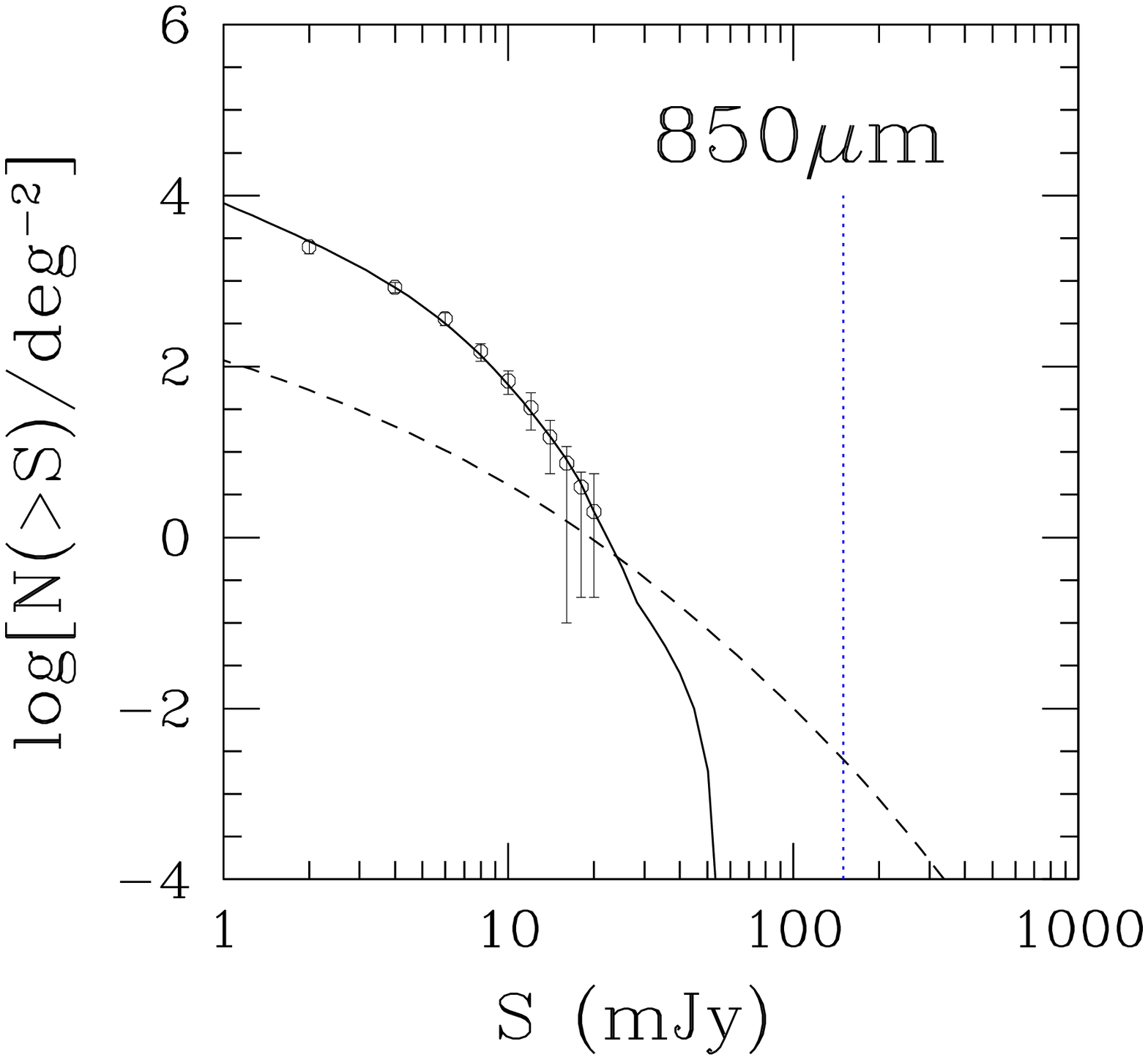} 
\caption{Source counts from \citet{coppin2006} plotted with a solid line showing
source count models of high-$z$
protospheroids from
\citet{granato2004}, rescaled to fit the source counts of \citet{coppin2006}. The dashed
line shows the prediction for strongly lensed protospheroids from \cite{negrello2007}. 
The vertical line shows the 5\,$\sigma$ depth of SASSy. The \cite{negrello2007} model predicts
that on the order of 40 strongly lensed protospheroids will be detected in the ``All-Sky''
phase.} 
\label{fig:lenses}
\end{figure} 

The high redshift luminous sub-mm galaxies that SASSy will detect are highly likely to be strongly
gravitationally lensed and SASSy offers a unique survey for rare high luminosity strong lenses.
The statistics of these lensing sources are a powerful probe of the geometry
of the Universe (e.g.~\citealt{gott1989}; \citealt{fukugita1992}).  This is illustrated in
Fig.~\ref{fig:lenses}, which shows the \citet{granato2004} model scaled to the source counts of
\citet{coppin2006}.  SASSy will detect on the order of 40 strongly
lensed (i.e.~gravitational amplification $\ge 2$) proto-spheroidal galaxies in the ``All-Sky'' Survey
phase \citep{negrello2007}. SASSy is thus ideally placed to find this  undiscovered population of bright lensed
sub-millimetre galaxies and  to use them to place new constraints on cosmological parameters such as
$\Omega_{\Lambda}$.

High redshift sub-mm galaxies will form an important component of SASSy, but we expect that the majority
of extragalactic sources detected by SASSy will be bright flat-spectrum blazars (otherwise known as BL
Lac objects or blazars). According to the predictions of \citet{negrello2007} and \citet{dezotti2005} at
a limiting flux level of 150 mJy/beam we will detect on the order of 0.054 blazars deg$^{-2}$, which
evaluates to 200-300 bright blazars in the GP-Wide and Pole-to-Pole strips. Recent wide-field MAMBO
observations have uncovered a surprising overdensity of millimetre bright blazars \citep{voss2006} and
SASSy will confirm the number counts of these bright blazars. Sub-mm bright blazars are excellent
candidates for ALMA phase calibrators due to their high brightness temperatures and consequent small
emitting areas \cite{holdaway2004}. The 20\,000 square degrees of the  "All-Sky" phase of SASSy will
contain  on the order of a thousand new potential phase calibrators for ALMA.

Finally, SASSy will also search for the  existence of cold local galaxies that were not
detected by IRAS. Sub-millimetre surveys of the local Universe  have so far mainly been
based upon the IRAS point source catalogue (e.g.~SLUGS; \citealt{dunne2000}) or on HI
catalogues (e.g.~SINGS; \citealt{kennicutt2003}). However, recent searches for Type Ia
supernova hosts have been uncovering  galaxies dominated by cold $\sim$20 K dust,and with
high $L_{850}$ at $z$=0.5 \citep{farrah2004}. Similar galaxies at $z$=0.1 would have
$F_{60}\sim 70$mJy, and so would not appear in any of the IRAS galaxy catalogues, but they
would have 850$\mu$m fluxes detectable by SASSy. Are there populations of cold,
ultraluminous galaxies in the local (i.e.\ $z\sim$0.1) Universe? The only way to determine
this is with a large area, shallow survey such as SASSy.

\subsection{Legacy value}

The wide area coverage and unbiased nature of SASSy will not only be instrumental in addressing the
Galactic and Extragalactic science goals outlined above, but will also be of significant
value to the wider astronomical community. IRAS shows the enduring value of wide area surveys in new
wavelength ranges; the IRAS mission has generated over 40\,000 citations since its launch
over 20 years ago and the rate of these citations has remained almost constant over the years,
reflecting the strong legacy value of IRAS data to the entire astronomical community. SASSy
aims to fulfil a similar goal in the sub-millimetre.

SASSy will act as a finder chart of bright sub-mm sources for ALMA and future instruments,
will provide target lists of dense galactic cores and luminous galaxies for future study
with infrared, radio and millimetre facilities, will supply catalogues of compact and
point-like sub-mm sources that could be used as flux and pointing calibrators  by
millimetre and sub-millimetre telescopes worldwide (including CCAT, LMT, ALMA and
Planck Surveyor). SASSy will also supply a map of the cold dust emission at small scales
over a large fraction of the sky and, as such, will be a valuable addition to the all-sky
dust extinction maps that the GAIA mission will produce and a valuable augmentation to the
Virtual Observatory at sub-mm wavelengths. It is the hope of the SASSy Consortium that data
from SASSy will be used as a lasting legacy well beyond the lifetime of the JCMT.


\section{Survey Timeline \& Data Products}

SASSy will commence with the rest of the JCMT Legacy Survey programme following the
successful commissioning of SCUBA-2, which is currently in the final stages of assembly for
delivery to the telescope in late 2007. Full survey operations are expected to commence in
early 2008 after the installation of the full complement of SCUBA-2 sub-arrays. We thus
expect to complete the pilot phase of SASSy, along with the remaining JCMT Legacy Surveys,
mid-way through 2010. The JCMT Board has agreed a proprietary period of one
year following the acquisition of the last JCMT Legacy Survey data and so the raw and
reduced data will nominally be released in 2011. The data products returned from SASSy will take the form of compact and point source
catalogues at 850 $\mu$m and image tiles. To achieve our science goals we will also need to
discriminate between galactic and extragalactic 850 $\mu$m emission, particularly at high
galactic latitudes.For this we will use existing archival CO and HI surveys (e.g. the International
Galactic Plane Survey\footnote{\texttt{http://www.ras.ucalgary.ca/IGPS/}}) to correlate the 850 $\mu$m
emission against, obtaining new data at high latitudes where necessary.

However, our goal is to release data from SASSy in advance of this date, available manpower permitting.
As already mentioned in the Introduction, SASSy offers the crucial opportunity to provide a
sub-millimetre survey catalogue before the final Herschel Announcement of Opportunity. We will release
our data in advance of this Announcement so that the wider community can fully exploit the legacy of
SASSy with Herschel. In addition, it may be possible to complete the pilot phase of SASSy early as we
will be observing in a fully flexibly-scheduled mode. The Mauna Kea weather statistics over the last
several years suggests that 15--20\% of time falls within the Grade 4 weather band. Assuming 300
observing nights per year, this results in $\sim$500 hours of Grade 4 observing time per year. If the
pressure upon the Grade 4 weather band is low (for example, following the retirement of the low
frequency single-pixel receivers on JCMT) then the pilot phase of SASSy may  be completed well within 2
years

\end{document}